# Phonon-assisted oscillatory exciton dynamics in monolayer MoSe$_2$

Colin M. Chow[1]*, Hongyi Yu[2]*, Aaron M. Jones[1], John R. Schaibley[3], Michael Koehler[4], David G. Mandrus[4,5], R. Merlin[6], Wang Yao[2]†, Xiaodong Xu[1,7]†

[1]Department of Physics, University of Washington, Seattle, Washington 98195, USA.

[2]Department of Physics and Center of Theoretical and Computational Physics, University of Hong Kong, Hong Kong, China.

[3]Department of Physics, University of Arizona, Tuscon, Arizona 85721, USA.

[4]Department of Materials Science and Engineering, University of Tennessee, Knoxville, Tennessee 37996, USA.

[5]Materials Science and Technology Division, Oak Ridge National Laboratory, Oak Ridge, Tennessee 37831, USA.

[6]Center for Photonics and Multiscale Nanomaterials and Department of Physics, University of Michigan, Ann Arbor, MI 48109, USA.

[7]Department of Materials Science and Engineering, University of Washington, Seattle, Washington 98195, USA.

*These authors contributed equally to this work.

†Correspondence: Wang Yao (Email: wangyao@hku.hk; Phone: 852-2219-4809; Address: Chong Yuet Ming Physics Building, The University of Hong Kong, Pokfulam Road, Hong Kong), Xiaodong Xu (Email: xuxd@uw.edu; Phone: 206-543-8444; Address: 3910 15th Ave N. E., Seattle, WA 98195)

**ABSTRACT**

**In monolayer semiconductor transition metal dichalcogenides, the exciton-phonon interaction strongly affects the photocarrier dynamics. Here, we report on an unusual oscillatory enhancement of the neutral exciton photoluminescence with the excitation laser frequency in monolayer MoSe$_2$. The frequency of oscillation matches that of the M-point longitudinal acoustic phonon, LA(M), suggesting the significance of zone-edge acoustic phonons and hence the deformation potential in exciton-phonon coupling in MoSe$_2$. Moreover, oscillatory behavior is observed in the steady-state emission linewidth and in time-resolved photoluminescence excitation data, which reveals variation with excitation energy in the exciton lifetime. These results clearly expose the key role played by phonons in the exciton formation and relaxation dynamics of two-dimensional van der Waals semiconductors.**

## INTRODUCTION

The electron-phonon interaction in solid state systems plays a major role in carrier dynamics[1], particularly in the relaxation of photoexcited carriers in semiconductor nanostructures, such as quantum wells[2], quantum wires[3], and quantum dots[4]. Monolayer semiconducting transition metal dichalcogenides (TMDs) have attracted much interest lately due to intriguing two-dimensional (2D) exciton physics, especially relating to their valley degrees of freedom[5,6]. In addition, reduced Coulomb screening in the 2D limit leads to nonhydrogenic exciton series[7–9] and strong many-body exciton physics[10,11]. Recently, signatures of a strong exciton-phonon interaction have been observed[12,13], such as the preservation of valley coherence in double-resonant Raman scattering[14], trion to exciton luminescence upconversion in monolayer WSe$_2$ assisted by $A_1'$ phonons[15], and exciton enhanced anti-Stokes shifts in few layer MoTe$_2$[16]. Despite a few theoretical proposals on the role of optical phonons in exciton dynamics[17–19], and several experimental studies on phonon-limited exciton relaxation[20–22], the details behind how and which phonons impact metrics such as the formation and relaxation of excitons remains largely unexplored. This knowledge is important for interpreting a wide range of 2D exciton phenomena and for exploring the potential of exciton-based 2D optoelectronics.

In this work, we investigate the role of exciton-phonon interaction in exciton dynamics using the model system of monolayer $MoSe_2$ (Fig. 1a). Performing photoluminescence excitation (PLE) spectroscopy, we observe that the neutral exciton PL intensity, as well as its linewidth, oscillates as a function of excitation energy with a period matching that of the longitudinal acoustic phonon at the M point, LA(M). Nested within the oscillations are fine structures, with linewidths one order of magnitude smaller than that of ordinary PL, originating from resonant Raman scattering. Analysis of the emission lineshape of the neutral exciton reveals that the oscillatory behaviour also presents in the homogeneous linewidth. Moreover, time-resolved PLE shows that exciton dynamics varies with respect to excitation energy, where shorter emission lifetime is measured for off-phonon-resonance excitation. This might due to the elevated lattice temperature arising from long-wavelength (small $k$-vector) acoustic phonons, which enhances radiative exciton recombination. Our results show that acoustic LA(M) phonons play an important role in electron-phonon coupling and hot-carrier cooling in monolayer $MoSe_2$, and also suggest the involvement of intermediate indirect excitonic states (with Q-valley electrons) in the formation of K-valley excitons.

**RESULTS AND DISCUSSIONS**

In our steady-state measurement, we detect PL at 5 K while scanning the excitation energy of a continuous wave (CW) laser, i.e., PLE spectroscopy (see Method Section for experimental details). The PLE intensity plot of Fig. 1b shows excitonic emission energies as a function of laser excitation. Two luminescence peaks are identified[23]: the neutral A exciton ($X^0$), centered at 1.650 eV, and the negative trion ($X^-$), centered at 1.624 eV. Evidently, the intensity of $X^0$ emission oscillates as a function of excitation laser frequency, while the behavior of $X^-$ is monotonic. Fig. 1c shows PL spectra taken at the excitation energies 1.699 eV (red) and 1.686 eV (green), which exemplify the contrasting excitation energy dependencies of $X^o$ and $X^-$ PL. Within our laser scan range, which has a high-energy limit of 1.77 eV (700 nm), five equally spaced regions of luminescence enhancement, indicated by the white arrows, can be seen in $X^0$, with an average energy separation of 18.5 meV. Such oscillations of $X^0$ emission intensity in PLE, first reported in CdS[24] for longitudinal optical phonons, is the hallmark of resonant excitation of phonon modes.

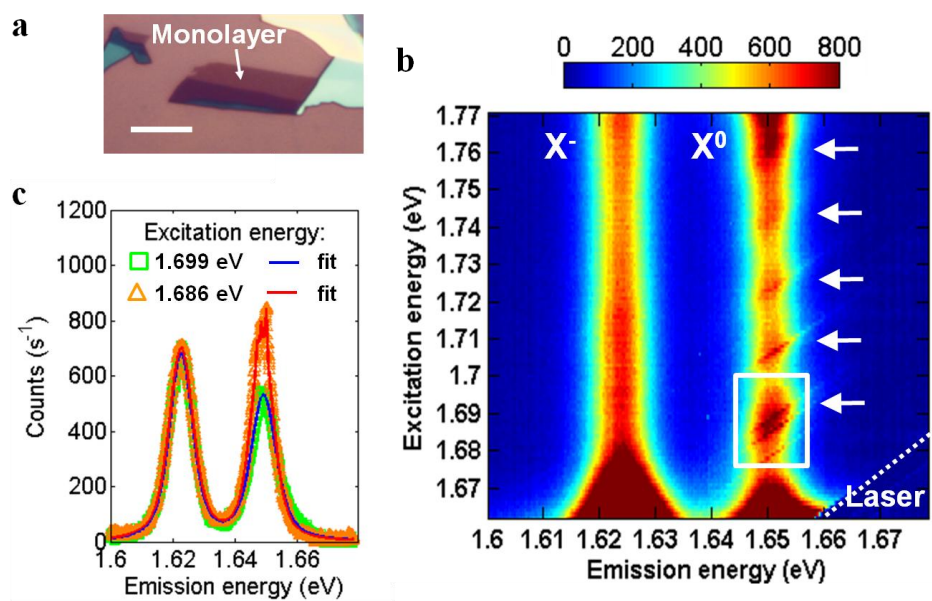

**Fig. 1 | Oscillatory enhancements of neutral exciton PL intensity in monolayer $MoSe_2$. a** Optical micrograph of an $MoSe_2$ monolayer on 285-nm-thick $SiO_2$ on silicon. Scale bar: 10 μm. **b** PLE intensity map of the monolayer shown in **a**, indicating neutral exciton ($X^0$) and trion ($X^-$) emission centered at 1.650 eV and 1.624 eV, respectively. Arrows indicate

regions of PL enhancement. Colour bar: counts per second. **c** PL spectra at two distinct excitation energies showing the variation of $X^0$ (but not $X^-$) PL with excitation energy.

A closer look at Fig. 1b shows that each PLE resonance region contains several narrow peaks. Fig. 2a offers an expanded view of the spectral regime highlighted by the white square in Fig. 1b. Three narrow lines shift in parallel with the excitation laser detuning, implying a Raman scattering origin of these lines. Their sub-meV linewidths are consistent with "conventional" Raman spectra measured on a different monolayer $MoSe_2$ sample (Supplementary Discussion), as well as with those reported in the literature[25–27]. The combined spectral features of PL emission and Raman scattering give rise to the overall emission spectrum, as shown in the example of Fig. 2b. As with earlier studies in monolayer $WSe_2$[14], on top of the spectrally broad features (conventional $X^0$ PL) sits a narrow peak arising from resonant Raman scattering. The intensity of broad $X^0$ PL changes gradually with excitation energy, resulting in a rising PLE background on both ends of the scan range, as apparent in Fig. 2c. This observation has been reported[28] and is most probably due to increased absorption near excitonic resonances, e.g., 1s excitonic state (1.650 eV) on the low energy side and 2s (1.830 eV) on the high energy side.

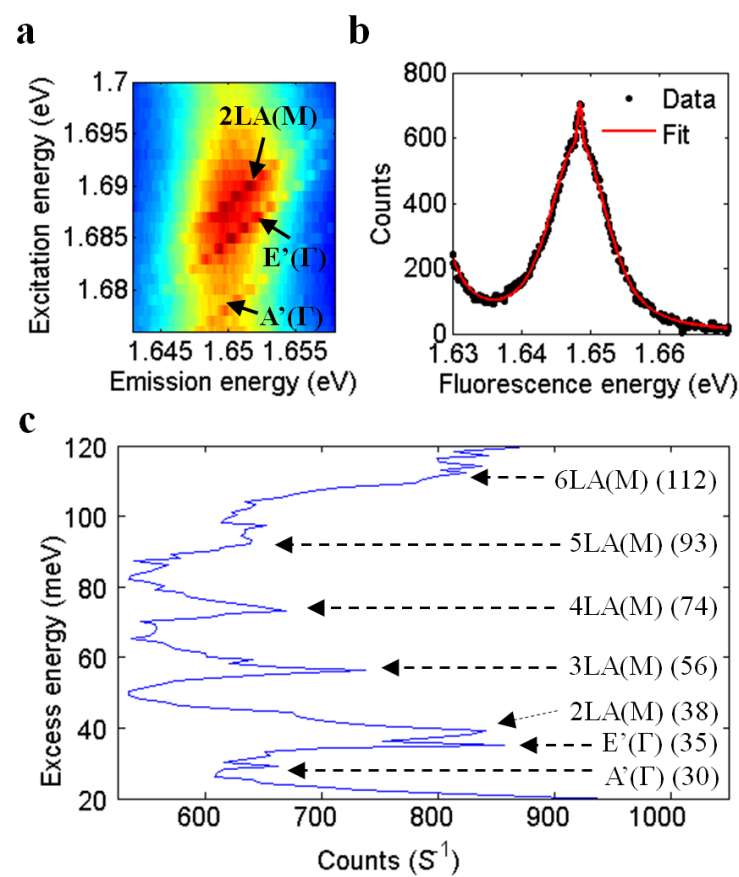

**Fig. 2 | Phonon peaks in monolayer $MoSe_2$ PLE spectra. a** Magnified view of the region enclosed by the white square in Fig. 1b, showing narrow resonance peaks with corresponding phonon modes indicated. **b** PL spectrum for 1.678 eV excitation, showing a narrow resonance associated with the $A'_1$ phonon superimposed on the broader $X^0$ emission. **c** Vertical line cut of the PLE map at the $X^0$ resonance, plotted in terms of excess energy, defined as the excitation energy subtracted by the $X^0$ resonance. Selected phonon enhanced peaks are labeled with excesses energy in parenthesis (in meV, with 1 meV uncertainty), along with possible assignments.

Now we turn to the assignment of the observed phonon modes. The dominant feature in the $X^0$ PLE is the average oscillation period of 18.5 meV. From recent studies of Raman scattering on monolayer $MoSe_2$[25,27], this period matches that of the M-point longitudinal acoustic phonon, LA(M). Fig. 3a shows the locations of M points in the Brillouin zone. According to *ab initio* calculations reported on monolayer $MoS_2$ and $WS_2$[29,30], the electron-phonon interaction strength is largest for LA phonons in the vicinity of the M points. Since monolayer $MoSe_2$ is structurally similar to $MoS_2$ and WS$_2$, we expect mode specific characteristics of electron-phonon coupling to qualitatively resemble those of these compounds[30]. Therefore, we assign the oscillation in PLE as overtones (harmonic series) of the LA(M). This assignment is corroborated by plotting the PL intensity at the neutral exciton resonance as a function of excess energy (Fig. 2c), defined as the photon energy difference between the excitation laser and the $X^0$ resonance (1.650 eV). Compared to the phonon assignment in Ref. 27, we identify the 38-meV peak as the resonant Raman scattering of the 2LA(M) mode. The other higher overtones from 3LA(M) to 6LA(M) are also identified and indicated in Fig. 2c with their respective excess energies. Likewise, we assign the 30-meV and 35-meV peaks to $A'_1(\Gamma)$ and $E'(\Gamma)$, respectively. Additional higher-order phonon mode assignments can be found in the Supplementary Discussion. Similar PLE features with identical phonon modes are also observed in a second sample (Supplementary Discussion). Results in Fig 1 and 2 suggest that the LA(M) mode plays a dominant role in hot exciton relaxation in monolayer MoSe$_2$, giving rise to periodical modulation of PL intensity of the A exciton as a function of excitation photon energy. The importance of LA(M) phonons in exciton dynamics is perhaps ubiquitous in TMDs based on recent studies in few-layer MoS$_2$[31] and WS$_2$[32].

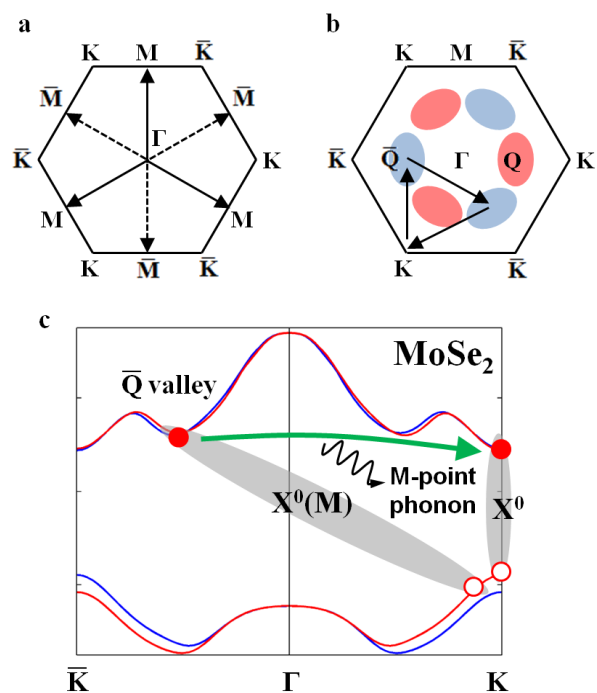

**Fig. 3 | Scattering between K-valley electrons and LA(M) phonons. a** Wavevectors of M-point phonons in a hexagonal Brillouin zone. **b** Phonon mediated transitions of an electron between K and $\bar{Q}$ valleys within the Brillouin zone. Here, three M-phonons are involved, with zero net change of electron *k*-vector. Blue- and red-shaded regions correspond to $\bar{Q}$ and Q valleys, respectively. **c** Lowest conduction band and top valence band of monolayer MoSe$_2$, with the $\bar{Q}$ valley indicated. Upon the LA(M) phonon mediated intervalley scattering of an electron between $\bar{Q}$ and K valleys, the optically dark indirect exciton $X^0(M)$ can be interconverted with the optically bright exciton $X^0$. Red and blue curves correspond to different spins.

As required by momentum conservation during the Raman process, the total phonon wavevector of each phonon enhanced PLE peak indicated in Fig. 2c must be zero. This requirement is easily fulfilled by $A'_1(\Gamma)$ and $E'(\Gamma)$ modes, but not the fundamental LA(M) mode. However, for the LA(M) overtones, the requirement can be satisfied through the following scenarios. In the case of 2LA(M), a combination of M and $\bar{M}$ wavevectors conserves momentum. In 3LA(M), this requirement is met following the scheme shown in Fig. 3b, where an equilateral triangle is formed by three M-vectors, also resulting in a zero vector sum. In monolayer MoSe$_2$, in addition to the K and $\bar{K}$ valleys (band edges), the conduction band also has Q and $\bar{Q}$ valleys located close to halfway between $\Gamma$ and K/$\bar{K}$ points. The momentum carried by an LA(M) phonon therefore matches the momentum separation between the K/$\bar{K}$ and $\bar{Q}$/Q valleys (Fig. 3a and b). Thus, following the involvement of M-point phonons, conservation of momentum stipulates that the electron be scattered between K- and $\bar{Q}$-valleys; see Fig. 3c. In other words, phonon-assisted scattering occurs between the optically bright exciton $X^0$ with both the electron and hole in K valley, and the optically dark indirect exciton $X^0$(M) with an electron in $\bar{Q}$ and a hole in K valley (Fig. 3c). Here, $X^0$(M) can be a virtual intermediate state such that its energy is not required to match that of $X^0$ + LA(M). Besides, despite the estimated 0.2-eV[33] electron band energy difference between K and Q valleys, the larger effective mass of the Q-valley[34] might result in a larger exciton binding energy of $X^0$(M) than that of $X^0$. This binding energy difference can partially cancel the electron band energy difference between Q and K valleys, leading to a smaller energy separation between $X^0$(M) and $X^0$, which enhances the role of $X^0$(M) as an intermediate state. A seemingly related intervalley exciton-phonon scattering is proposed to explain the strong 2LA(M) peak in excitation-dependent Raman spectroscopy of $WS_2$[32], although the excitation therein is well above band edge and involves higher lying conduction bands.

We note that while oscillations due to phonon resonance feature prominently in the $X^0$ transition, the $X^-$ emission intensity is relatively constant, except for excitation below 1.68 eV, close to the $X^0$ resonance of 1.65 eV. The lack of oscillatory enhancement in $X^-$ is possibly due to its distinct radiative properties compared to $X^0$, together with the availability of multiple formation pathways[35] (e.g. via the exciton-electron interaction following exciton relaxation[36]). For $X^0$, only those inside the light cone ($k \leq \omega_{X^0}/c$) can radiate. In contrast, $X^-$ with a much larger range of momentum can radiate due to the electron recoil effect[23]. As a result, $X^0$ PL intensity depends strongly on its momentum distribution, as determined by the resonance condition of the excitation energy, while such dependency is weak for $X^-$. Moreover, the $X^-$ formation process is largely independent of the excitation energy, because even for excitation away from the phonon resonances, optically dark excitons can still be generated at large momentum (outside the light cone), which can interact with electrons to form trions. The lack of sensitivity to the excitation energy in both trion formation and relaxation processes diminishes any oscillatory features in the $X^-$ emission. A more detailed discussion can be found in the Supplementary Discussion.

Aside from spectral information, the PLE map also offers insights into the exciton dynamics. From the fit to the spectrum taken at each excitation energy, we found that both $X^-$ and $X^0$ lineshapes are well-described by Voigt profiles, from which one can infer the homogeneous linewidths of the excitonic transitions, as well as the widths of the Gaussian-broadened spectral distributions of their resonances (Supplementary Discussion). The latter is associated with the inhomogeneous broadening of the excitonic transitions. Oscillatory behavior is found in the homogeneous linewidth, $\gamma_0$, of $X^0$, which is smaller for excitation resonant with phonon harmonics (Fig. 4a). Its inhomogeneous (Gaussian) width remains relatively constant, consistent with the expectation that inhomogeneous broadening should depend only weakly on excitation energy. $\gamma_0$ is associated with the coherence lifetime of the exciton, and is given by $\gamma_0 = \gamma/2 + \Gamma$, where $\gamma$ is the inverse of exciton population lifetime and $\Gamma$ the pure dephasing rate. Since $\Gamma$ is proportional to the rate of dephasing processes such as exciton-phonon scattering[22], it is reasonable to assume that the oscillations in $\gamma_0$ is largely due to the creation of long-wavelength phonons during the relaxation of excited (hot) excitons. Our analysis on time-resolved PLE data (more details discussed below and in the Supplementary Discussion) seems to support this interpretation.

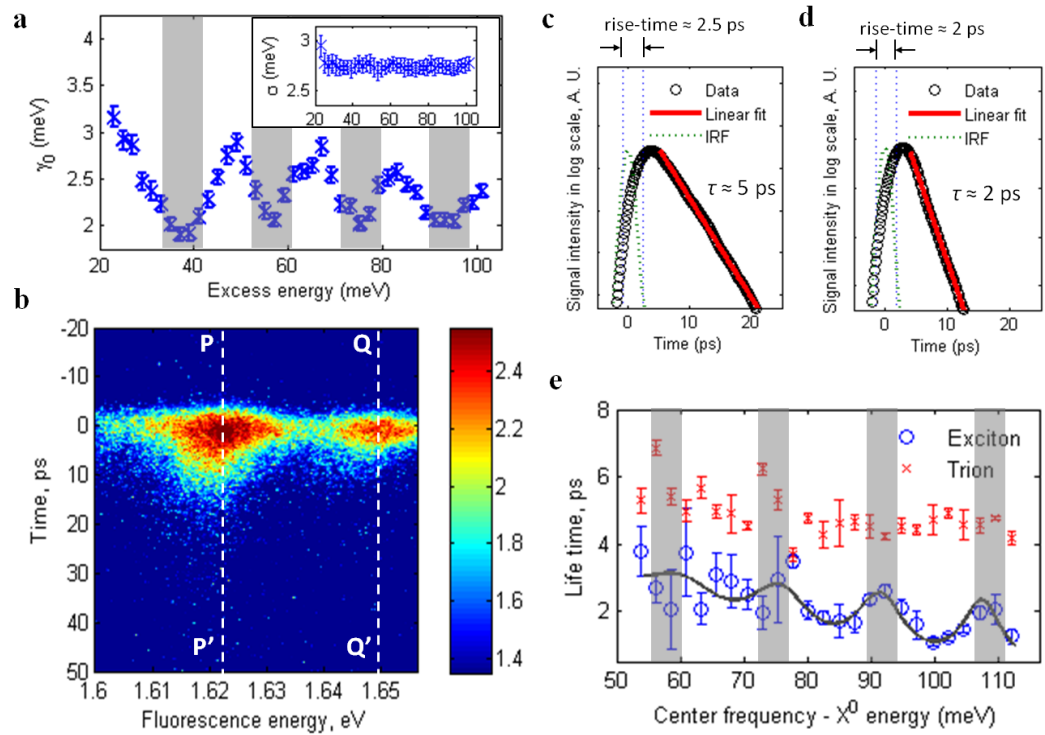

**Fig. 4 | LA(M) phonon induced modulation of exciton dynamics in monolayer $MoSe_2$. a** Oscillations in the best-fit homogeneous linewidth, $\gamma_0$, of the $X^0$ resonance with the excess energy as defined in Fig. 2c. Error bars represent 99% confidence intervals of the fits. Inset: Width of inhomogeneous (Gaussian) broadening, $\sigma$. **b** An example of raw streak camera data for excitation centered at 1.732 eV. The trion ($X^-$) and neutral exciton ($X^0$) resonances are marked by dashed lines P – P' and Q – Q', respectively. Colour bar: signal intensity in log-scale, arbitrary unit. **c, d** Time-traces of the $X^-$ and $X^0$ resonances, respectively, after a 2D Gaussian filter and deconvolution (Supplementary Discussion), in log-scale. The 10% – 90% rise-times are indicated, along with the best-fit exponential decay lifetimes, $\tau$. Circles and red solid lines represent data points and linear fits to the exponential decay, respectively. **e** Lifetimes, $\tau$, of $X^-$ (crosses) and $X^0$ (circles) extracted from a series of streak camera measurements with varying excitation energy. In the x-axis, similar to how the excess energy is defined, a constant $X^0$ energy (1.650 eV) is subtracted from the excitation center frequency. Error bars represent standard deviations, while the dashed line serves as guide to the eye. Shaded regions indicate phonon resonances as obtained from Fig. 1b.

To explore the phonon-assisted dynamics directly, we measure time-resolved emission of $X^0$ and $X^-$ with a streak camera. Fig. 4b presents an example of the measured spectra with the pulsed excitation centered at 1.732 eV. The time evolution of the emission is characterized by a rapid onset within a few picoseconds, followed by an exponential decay. This is shown by extracting time traces along P – P' and Q – Q' from Fig. 4b for $X^-$ and $X^0$ and plotted in Figs. 4c and d, respectively. Here, a two-dimensional Gaussian filter is applied to the raw data, followed by deconvolution with Tikhonov regularization (see Supplementary Discussion), resulting in smooth temporal profiles from which the 10% – 90% rise time and lifetime ($\tau$) can be estimated. By stepping the centre frequency of the pulses, excitation energy dependent rise times and lifetimes are obtained and the latter is plotted in Figs. 4e, where shaded regions indicate energies of phonon-enhanced PL seen in the PLE map in Fig. 1b.

Although fluctuation with the excitation energy is apparent in the rise times of $X^0$ and $X^-$ emission (Supplementary Discussion), due to the measurement noise and the resolution of the streak camera, we are unable to observe an unambiguous systematic variation with the excitation energy. A further study with improved experimental method and cleaner data is needed to determine whether the rise times oscillate in accordance to the phonon modes seen in Fig. 1b. Nonetheless, in Fig. 4e showing the emission lifetimes, despite the noise caused by excitation pulse leak-through at energies below 1.73 eV, the remaining data shows clear oscillations of the $X^0$ emission lifetime, i. e., the exciton relaxation dynamics is affected by resonant excitation of the LA(M) phonon mode. To understand this effect, we propose a framework using rate equations to model phonon-assisted interconversions between excitons inside and outside the light cone (Supplementary Discussion). In brief, a thermalized population of excitons both inside and outside the light cone is formed shortly after excitation. The bright (inside the light cone) excitons quickly recombine[37,38], leaving behind the dark excitons, which are then scattered into the light cone by phonons (Fig. 5a) and recombine at a later time, producing the observed exponential decay. When the excitation is off-resonant, excitonic relaxation results in the emission of many long-wavelength phonons (Fig. 5b), forming a phonon bath that increases the scattering rate. On the other hand, on-resonance excitation (Fig. 5c) produces only a small number of LA(M) phonons which are ineffective in the aforementioned scattering process due to momentum mismatch. This picture is consistent with the oscillations of $\gamma_0$ shown in Fig. 4a, i.e., the creation of long-wavelength phonon bath during off-resonance excitation increases exciton-phonon scattering and manifest as a broadening of the $X^0$ homogeneous linewidth.

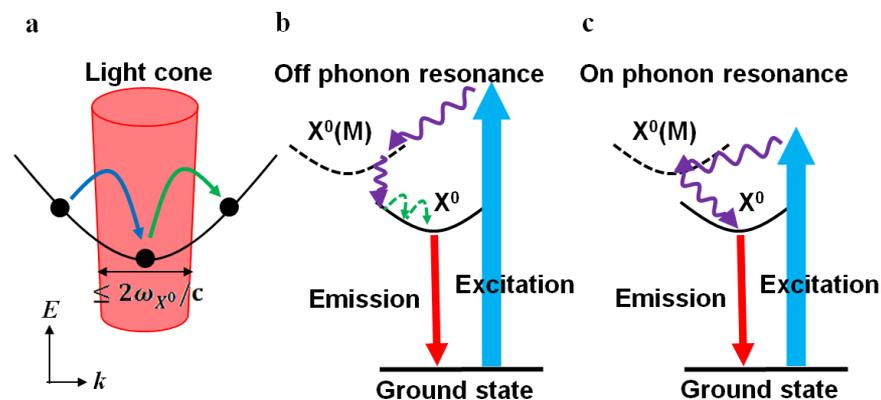

**Fig. 5 | Exciton-phonon scattering processes with off/on phonon resonance excitation. a** Schematic showing inward and outward scattering of the excitons across the light cone due to long-wavelength phonons. **b** Energy level scheme showing optical excitation (blue arrow) away from a phonon resonance, intervalley relaxations accompanied by the emission of LA(M) phonons (purple wavy arrows) and the eventual fluorescence-producing radiative decay of $X^0$ (red arrow). The laser generates $X^0$ with finite kinetic energies following the emission of LA(M) phonons. $X^0$ then relaxes to the minimum energy by emitting long wavelength phonons (green curved arrows). **c** On-phonon-resonance excitation where only a small number of LA(M) phonons are emitted in the ensuing relaxation.

To summarize, we observed excitation energy dependent oscillatory behavior of $X^0$ luminescence and dynamics, which largely stems from resonant excitation of LA(M) phonon modes in monolayer $MoSe_2$. Remarkably, it is the phonon at the Brillouin zone edge (M-point) that dominates the multiple-phonon scattering during hot exciton relaxation. This suggests the involvement of an intermediate $X^0(M)$ state and may present an opportunity to investigate the dynamics of intervalley excitonic transitions in 2D semiconductors. The prevailing involvement of acoustic phonons, rather than of optical phonons, is in agreement with DFT calculations[29,30]. This implies that the deformation potential coupling dominates over Fröhlich coupling in our samples, although the converse is normally expected, as reported in CdS[24]. Further studies are needed to fully elucidate the process of carrier-phonon interactions in semiconducting TMDs, especially with regard to material- and phonon-mode-specific details. Nonetheless, our results can potentially be exploited to understand exciton related physics and design optoelectronics applications based on 2D semiconductors.

**METHODS**

Bulk $MoSe_2$ crystals are synthesized using chemical vapor transport technique with iodine as transport agent. To obtain monolayer $MoSe_2$ samples, thin $MoSe_2$ flakes are mechanically exfoliated from bulk crystals onto 285-nm-thick $SiO_2$ thermally grown on Si wafers. Monolayers are first identified visually by their optical contrast under a microscope and then confirmed by measuring their thickness (~0.7 nm) using atomic force microscopy. Throughout the experiments, the samples are maintained at a temperature of 5 K in a cold finger cryostat.

All optical studies are made in reflection geometry, where the incident beam and fluorescence traverse the same microscope objective mounted on a micrometer stage assembly. The use of a high-power microscope objective with cover-slip correction results in a focal spot size of less than 1 μm in diameter. For PLE measurements, a Ti-Sapphire tunable CW laser (Solstis from M-Squared Lasers, LTD) is used to produce the excitation beam. The incident power is held at 20 μW, below the PL saturation threshold of about 100 μW. The excitation frequency is scanned at a rate of 5 seconds per step and with a step size of 1 meV (~ 0.42 nm at 721 nm). Rejection of the reflected excitation laser is accomplished first by an analyzer and then by spectral filtering. The spectral filter consists of a pair of achromatic doublets forming a 1:1 telescope, and a pair of gratings positioned at the outward conjugate focal planes. At the central focal plane, the spectral distribution of the fluorescence is mapped into spatial separation, and a graphite rod mounted on a translational stage is used to block the frequency component belonging to the excitation laser. Finally, fluorescence spectra are measured with a liquid-nitrogen-cooled CCD camera attached to the output port of a spectrometer (Princeton Acton SP2500). The final spectral resolution is 0.027 nm (~0.06 meV at 755 nm), limited by the pixel size of the CCD camera. Each spectrum is taken with an accumulation time of 5 seconds.

In time-resolved PLE measurements, the excitation beam is produced by a wavelength-tunable, mode-locked Ti-Sapphire laser with a pulse repetition rate of 80 MHz, and a pulse duration of 150 fs. The bandwidth of the output pulses, of about 17 meV, is unsuitable for resolving the oscillatory behavior, which is expected to have a period of about 18.5 meV. The pulse bandwidth is therefore reduced to about 5 meV with the aid of the spectral filter described above, only with the graphite rod now replaced by a slit. This introduces a chirp to the pulses, which is then compensated by passing the beam through a single mode fiber with a predetermined length. The fiber serves a secondary purpose of producing a clean Gaussian beam profile to help achieving the best beam spot on the sample. The central wavelength of the pulses is tuned with 1-nm steps (~2.4 meV at 721 nm), and the average power is held constant at 50 μW. This corresponds roughly to a photogenerated exciton density on the order of $10^{11}$ - $10^{12}$ cm$^{-2}$. A streak camera (Hamamatsu C10910-05) with a nominal resolution of 1 ps is used to register the time-resolved fluorescence spectra. The streak camera is operated at a moderate gain, optimized for the best signal-to-noise ratio, with a 100-s integration time for each spectrum.

Procedures for PL lineshape fitting, data post-processing and analysis, and further discussions can be found in the Supplementary Discussion accompanying the paper.

## DATA AVAILABILITY

Data that supports the findings of this study is available from the corresponding authors upon reasonable request.

## ACKNOWLEDGEMENTS

This work is mainly supported by the Department of Energy, Basic Energy Sciences, Materials Sciences and Engineering Division (DE-SC0008145 and SC0012509). H.Y. and W.Y. are supported by the Croucher Foundation (Croucher Innovation Award), HKU ORA, and the RGC and UGC of Hong Kong (HKU17305914P, AoE/P-04/08). M.K. and D.G.M. are supported by US DoE, BES, Materials Sciences and Engineering Division. RM is supported by the MRSEC Program of the NSF under Grant No. DMR-1120923. XX acknowledges a Cottrell Scholar Award, support from the State of Washington funded Clean Energy Institute, and Boeing Distinguished Professorship.

**COMPETING INTERESTS**

The authors declare no competing financial interests.

**AUTHOR CONTRIBUTIONS**

XX and WY conceived and supervised the experiments. CMC fabricated the sample and performed the measurements assisted by AMJ and JRS. CMC, XX, HY, WY, RM analyzed the data. MK and DGM provided and characterized the bulk $MoSe_2$. CMC, XX, WY, HY, RM wrote the paper. All authors discussed the results.

# Phonon-assisted oscillatory exciton dynamics in monolayer MoSe$_2$: Supplementary discussion

Colin M. Chow[1]*, Hongyi Yu[2]*, Aaron M. Jones[1], John R. Schaibley[3], Michael Koehler[4], David G. Mandrus[4,5], R. Merlin[6], Wang Yao[2†], Xiaodong Xu[1,7†]

This supplementary document is comprised of six sections: (i) PL lineshape fitting, (ii) exciton rise time and lifetime extraction, (iii) exciton and trion decay rates in time-resolved luminescence, (iv) exciton and trion steady-state dynamics, (v) data from a second monolayer MoSe$_2$ sample, and (vi) Raman spectrum from a third monolayer MoSe$_2$ sample. To simplify referencing, all labels for figures and tables in the Supplementary Discussion begin with the letter "S".

## I. PL lineshape fitting

The PL lineshapes of both neutral exciton (X$^0$) and trion (X$^-$) are found to be inhomogeneously broadened in such a way that the broadening is on the same order of magnitude as the natural linewidth. Therefore, neither Lorentzian nor Gaussian lineshapes adequately fit the spectra, as shown in Fig. S1a. The shortcoming of these lineshapes is most evident in the "wings" of the resonances, where the best-fit Lorentzian curve overestimates the signal while the Gaussian curve underestimates it. As mentioned in the main text, we fit the spectra with Voigt profiles, or the plasma dispersion functions. Explicitly, the Voigt profile, $V(\omega; \omega_0, \sigma, \gamma)$, is given by the convolution between a Gaussian curve, $G(\omega; \sigma)$, and a Lorentzian curve, $L(\omega; \gamma)$:

$$V(\omega; \omega_0, \sigma, \gamma_0) = A \int_{-\infty}^{\infty} G(\omega' - \omega_0; \sigma) L(\omega - \omega'; \gamma_0) d\omega',$$
$$G(\omega; \sigma) = \frac{1}{\sigma\sqrt{2\pi}} \exp(-\omega^2/2\sigma^2),$$
$$L(\omega; \gamma_0) = \frac{\gamma_0}{\pi(\omega^2 + \gamma_0^2)}.$$

Here, $A$ is a proportionality constant, $\omega_0$ and $\sigma$ the average (centre) and standard of deviation of the normally (Gaussian) distributed resonance, respectively, and $\gamma_0$ the half width at half maximum (HWHM) of the Lorentzian. For every PL spectrum, two separate sets of fitting parameters, $A$, $\omega_0$, $\sigma$, and $\gamma_0$ are needed for the X$^-$ and X$^0$ resonances. Fitting is accomplished by minimizing the squared error by performing multivariable nonlinear regression. Some spectra feature prominent narrow lines attributed to resonant Raman scattering. These are fitted manually with Gaussian curves. Since the linewidths of these narrow features are much less than that of X$^-$ and X$^0$, their presence has minimal impact on the accuracy of the Voigt profile fits. As an example, the full-spectrum fit of Fig. 2b in the main text is shown in Fig. S1a below, along with fits using Lorentzian and Gaussian curves. The excellent fits produced with Voigt profiles for both X$^-$ and X$^0$ resonances justify their use.

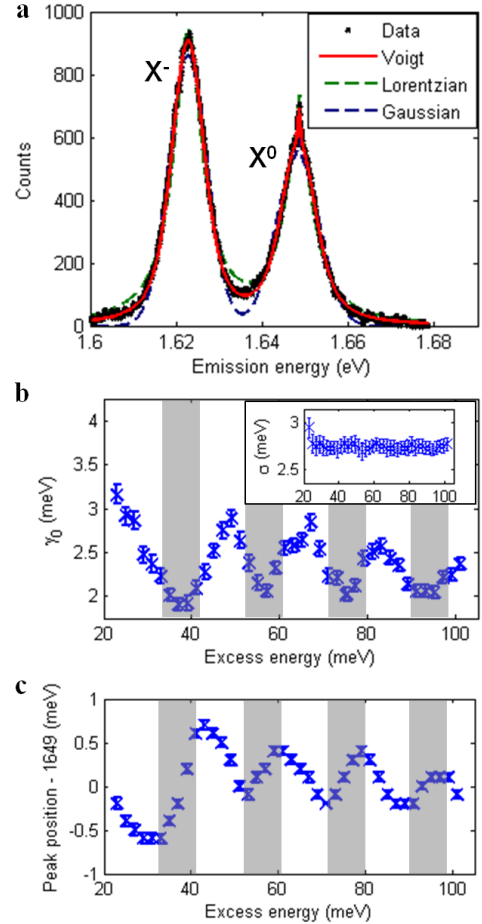

**Fig. S1** **a** Full spectrum of Fig. 2b in the main text showing lineshape fits with Voigt, Lorentzian and Gaussian profiles, for both X$^-$ and X$^0$ resonances. **b** Same as Fig. 4a in the main text, reproduced here for convenience: best-fit $\gamma_0$ of the X$^0$ resonance. **c** The centre of the X$^0$ resonance as a function of excess energy. Shaded regions indicate regions of enhanced luminescence obtained from Fig. 1b. Error bars represent 99% confidence intervals, defined as ±2.58 × standard error of fitted parameter.

[1]Department of Physics, University of Washington, Seattle, Washington 98195, USA. [2]Department of Physics and Center of Theoretical and Computational Physics, University of Hong Kong, Hong Kong, China. [3]Department of Physics, University of Arizona, Tuscon, Arizona 85721, USA. [4]Department of Materials Science and Engineering, University of Tennessee, Knoxville, Tennessee 37996, USA. [5]Materials Science and Technology Division, Oak Ridge National Laboratory, Oak Ridge, Tennessee 37831, USA. [6]Center for Photonics and Multiscale Nanomaterials and Department of Physics, University of Michigan, Ann Arbor, MI 48109, USA. [7]Department of Materials Science and Engineering, University of Washington, Seattle, Washington 98195, USA.
*These authors contributed equally to this work.
†Correspondence: Wang Yao (wangyao@hku.hk), Xiaodong Xu (xuxd@uw.edu)

The best-fit values of the fitting parameters are given in Table S1. Interestingly, the best-fit $\gamma_0$ and $\omega_0$ for $X^0$ vary in an oscillatory manner with respect to the excitation photon energy, whereas those of $X^-$ remain relatively constant. As shown in Fig. S1b, $\gamma_0$ of $X^0$ is reduced when the excitation laser is on a phonon resonance. Surprisingly, Fig. S1c reveals that $\omega_0$ also oscillates with a period corresponding to the LA(M) phonon mode. However, the oscillations are not in phase with that of $\gamma_0$ because here the rising edges coincide with the phonon resonances. Artifacts from fitting can be ruled out with further analysis of the fitting error, e. g. from the lack of corresponding oscillations in the residue. The amplitude of oscillations roughly equals that of $\gamma_0$, but much smaller than the average homogeneous linewidth. This might be related to the oscillations of $\gamma_0$, in which efficient hot exciton relaxation and the subsequent smaller phonon bath is conjectured to prolong radiative lifetime (Section IV), hence reducing $\gamma_0$, with on-phonon-resonance excitation. Nonetheless, the physical mechanism behind the modulation of exciton resonance seen here remains to be explored.

| Resonance | Parameter | Best-fit value (meV) |
|---|---|---|
| $X^-$ | $\gamma_0$ | 2.30 ± 0.05 |
| | $\sigma$ | 2.80 ± 0.05 |
| | $\omega_0$ | 1622.4 ± 0.1 |
| $X^0$ | $\gamma_0$ | Varies. See Fig. S1b |
| | $\sigma$ | 2.80 ± 0.05 |
| | $\omega_0$ | Varies. See Fig. S1c |

**Table S1** Best fit parameter values for both $X^-$ and $X^0$ transitions. With the exception of $\gamma_0$ and $\omega_0$ for the $X^0$ transition, all other parameters given here are found to vary only slightly with respect to the excitation energy, and without a clear oscillatory feature.

## II. Exciton rise time and lifetime extraction

In time-resolved PLE measurements, due to the finite response time of the streak camera, the output is the convolution of the time-dependent emission with the instrument response function, IRF. Since the timescale of MoSe$_2$ exciton dynamics is in the picosecond range, on the same order of magnitude as the nominal resolution (1 ps) of the streak camera, the measured time evolution of the emission will be noticeably stretched by about 1 ps. To obtain more accurate timing parameters, we perform deconvolution on the streak camera output with linear least-squares regression[1]. By scattering light from a mode-locked Ti-Sapphire laser with 150 fs pulse width into the streak camera, we measured a 1.5 ps full-width-at-half-maximum (FWHM) response. This is shown in Fig. S2a (dotted line) and taken as the IRF for deconvolution. To avoid overfitting and to improve numerical stability of the deconvolution protocol, Tikhonov regularization[1] is applied. With an appropriate choice of Tikhonov factor, the noise in the reconstructed temporal profile can be suppressed without affecting the overall response, as exemplified by the time traces of the trion emission, before (dashed line) and after (solid line) deconvolution, in Fig. S2a. The same deconvolution procedure is applied to all streak camera images taken at different excitation frequencies. Expectedly, we find that the rise times of $X^0$ and $X^-$ emission are reduced by roughly 1 ps following deconvolution, as compared with those extracted from the raw data. Nonetheless, their variations with respect to excitation frequency remain qualitatively similar with or without deconvolution.

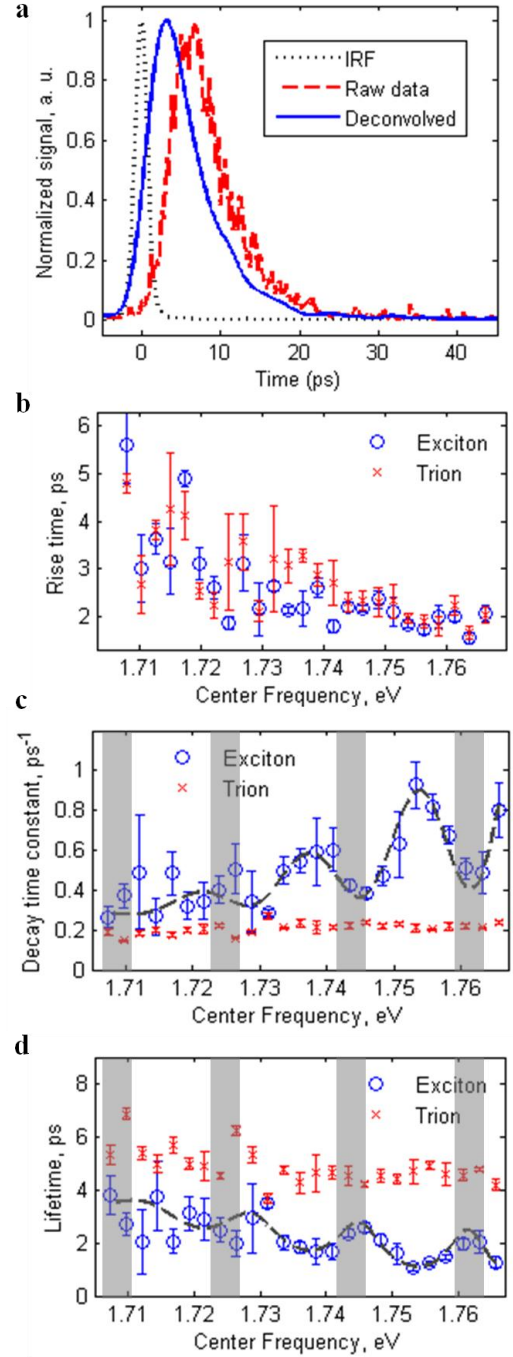

**Fig. S2** **a** Vertical time traces of trion emission from Fig. 4b. Red dashed curve is the raw streak camera measurement while blue solid curve the deconvolved data, with the IRF represented by the dotted line. **b** Excitation frequency-dependent rise times of $X^-$ (crosses) and $X^0$ (circles) emission extracted from a series of streak camera measurements in varying excitation energy. Error bars represent the standard deviations. **c** Excitation frequency-dependent decay rate, $\gamma$, of $X^-$ (crosses) and $X^0$ (circles). **d** Same data as in **c**, presented as the lifetime, $\tau$. Reproduced from Fig. 4e.

From the deconvolved data, we obtain the rise times and lifetimes of $X^0$ and $X^-$ with respect to excitation energy as presented in Fig. S2b – d. The rise time is simply defined as the time interval for which the rising edge of the luminescence moves from 10% to 90% of peak emission. To obtain the decay rate, a linear fit to the logarithm of the falling edge is applied, where the slope gives the decay time constant, $\gamma$. The extracted excitation dependent decay rates for the first sample is shown in Fig. S2c. To facilitate comparison with the rise time, the data in the main text is presented as the lifetime, $\tau = 1/\gamma$. For the conversion of the error bars from $\varepsilon(\gamma)$ to $\varepsilon(\tau)$, the first order approximation $\varepsilon(\tau) = \varepsilon(\gamma)/\bar{\gamma}^2$ is adopted, where $\bar{\gamma}$ denotes the mean value of $\gamma$. The result of the conversion is reproduced here in Fig. S2d.

## III. Exciton and trion decay rates in time-resolved luminescence

The exciton dispersion curve is shown in Fig. S3a, where only excitons inside the light cone ($k \leq \omega_{X^0}/c$) can radiatively recombine. To model the exciton decay process, we consider a simplified three-level system as shown in Fig. S3b. $|D\rangle$ represents the dark exciton state outside the light cone ($k > \omega_{X^0}/c$), with time dependent density $N_D(t)$, which nonradiatively decays to the vacuum state $|0\rangle$ with a rate $\Gamma_{nr}$. $|B\rangle$ is the bright exciton state inside the light cone ($k \leq \omega_{X^0}/c$), with time-dependent density $N_B(t)$, which decays to $|0\rangle$, both radiatively and nonradiatively, with a total rate $\Gamma_t = \Gamma_{nr} + \Gamma_r$. The nonradiative decay rate includes relaxation from excitons to trions. Scattering with other excitons, free carriers, impurities, and long-wavelength (small $k$) phonons can induce interconversion from $|D\rangle$ to $|B\rangle$ and vice versa, with the corresponding rates given by $\gamma_1$ and $\gamma_2$, respectively. Given the ~meV linewidth of excitons inside the light cone[2], interconversion from $|D\rangle$ to $|B\rangle$ can be induced by both phonon emission and absorption, whose coupling matrix elements are proportional to $\sqrt{N_\mathbf{k} + 1}$ and $\sqrt{N_\mathbf{k}}$, respectively, where $N_\mathbf{k}$ is the number of $\mathbf{k}$-vector phonon. Consequentially, $\gamma_1$ and $\gamma_2$ increase with $N_\mathbf{k}$.

$N_B(t)$ and $N_D(t)$ are governed by the coupled rate equations

$$\frac{d}{dt}N_B = -(\Gamma_t + \gamma_2)N_B + \gamma_1 N_D,$$
$$\frac{d}{dt}N_D = \gamma_2 N_B - (\Gamma_{nr} + \gamma_1)N_D.$$

Solving the above rate equations, we find

$$N_B(t) = Ae^{-x_f t} + Be^{-x_s t},$$
$$N_D(t) = A'e^{-x_f t} + B'e^{-x_s t},$$

where $A$, $B$, $A'$ and $B'$ are time-independent constants determined by the initial values $N_{B,D}(0)$. The PL emission rate is given by $\Gamma_r N_B(t)$, which has two decay time constants, $x_f$ and $x_s$. Under a low temperature we can take[3] $\Gamma_r \sim 3 \text{ ps}^{-1} \gg \gamma_1$ such that $(\Gamma_r + \gamma_2 - \gamma_1)^2 \gg 4\gamma_1\gamma_2$, the fast (slow) decay constant $x_f$ ($x_s$) is then given by

$$x_f = \frac{\Gamma_t + \gamma_2 + \Gamma_{nr} + \gamma_1}{2} + \sqrt{\left(\frac{\Gamma_t - \Gamma_{nr} + \gamma_2 - \gamma_1}{2}\right)^2 + \gamma_1\gamma_2} \approx \Gamma_t + \gamma_2.$$
$$x_s = \frac{\Gamma_t + \gamma_2 + \Gamma_{nr} + \gamma_1}{2} - \sqrt{\left(\frac{\Gamma_t - \Gamma_{nr} + \gamma_2 - \gamma_1}{2}\right)^2 + \gamma_1\gamma_2} \approx \Gamma_{nr} + \gamma_1.$$

Now $x_f$ corresponds to the total decay rate of the bright excitons inside the light cone, and $x_s$ that of the dark excitons outside the light cone.

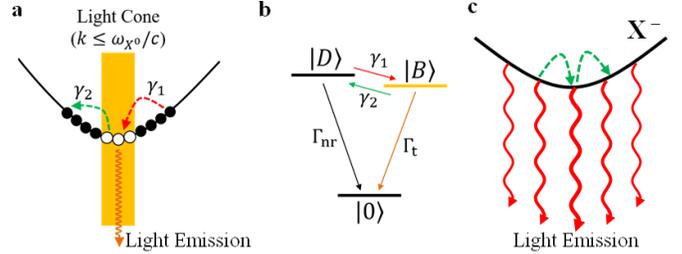

**Fig. S3** **a** Dispersion relation of neutral exciton, where the horizontal axis represents center-of-mass momentum. The light cone (yellow region) corresponds to center-of-mass wavevector, $k \leq \omega_{X^0}/c$, in which excitons are bright and can radiatively recombine (wavy arrow). Excitons outside the light cone are optically dark. Interconversion between bright and dark excitons (dashed arrows) can occur via scattering. **b** Three level model for neutral excitons, with relaxation rates indicated. $|D\rangle$: dark exciton, $|B\rangle$: bright exciton, $|0\rangle$: vacuum state. **c** Dispersion relation of the trion (solid black curve) and its radiative decay (red wavy arrows). Unlike the exciton, the radiative decay rate of the trion varies slowly with **k**, such that the scattering induced by long-wavelength phonons (green curved arrows) barely changes the overall decay rate. The thickness of the wavy arrows illustrates varying decay rate with respect to wavevector.

Theory has shown that bright excitons in monolayer TMDs have a very short radiative lifetime[3], as evidenced by the measured decay rate[2,4] of $x_f \sim 3 \text{ ps}^{-1}$. This value is significantly larger than the inverse of PL rise time (0.2 – 1 ps$^{-1}$) and decay rate (0.2 – 1 ps$^{-1}$) shown in Fig. S2b and c. Possibly, in our time-resolved measurements, by the time the recorded luminescence intensity reaches its maximum, the bright exciton density inside the light cone is depleted. The luminescence detected at a later time predominantly comes from scattering-induced dark-to-bright exciton conversion, whose decay rate is given by $x_s \approx \Gamma_{nr} + \gamma_1 \sim 0.2 - 1 \text{ ps}^{-1}$. The variation of PL decay time ($x_s^{-1}$) mainly comes from the modulation of the scattering rate, $\gamma_1$, which increases with the density of long-wavelength phonons. A related behavior due to the weak scattering rate at low temperature, termed "exciton-phonon relaxation bottleneck", has also been discussed in a recent paper[5].

In contrast to the exciton, trion has rather different radiative properties[3]. In particular, a trion with large momentum can radiatively decay, as the momentum conservation can be satisfied with the left-behind electron inherits the trion wavevector (the electron recoil effect[6]). The radiative decay rate varies slowly with the trion wavevector[3], such that scattering with long wavelength-phonons barely changes the overall decay rate of the trion (see Fig. S3c). Therefore, unlike the exciton, trion does not show lifetime oscillation with the excitation energy.

## IV. Exciton and trion steady-state dynamics

The contrasting behaviors of exciton and trion luminescence in PLE with excitation energy can be explained by their distinct radiative properties. For the excitons, momentum conservation allows only those inside the narrow light cone ($k \leq \omega_{X^0}/c \sim 10^{-3}$ Å$^{-1}$) to radiatively recombine. These bright excitons have an ultrafast radiative decay rate[3] of about 3 ps$^{-1}$. On the other hand, trions in a much larger momentum range can radiatively recombine, with a smoothly changing decay rate[3,6].

The PLE spectra are determined by the steady-state population distribution of excitons/trions. We expect the steady-state total exciton/trion population to depend weakly on the excitation energy because: (i) on- or off-phonon resonance affects only the kinetic energy of the generated exciton, while the generation rate is nearly unaffected as it is determined by the LA(M) phonon emission rate; (ii) most of the trions are the outcome of subsequent relaxation from excitons[7] which is insensitive to the exciton kinetic energy (rather, trion formation rate is mostly determined by phonon emission rates of $A_1'$ and other optical phonons[8]).

Nonetheless, the excitation energy strongly affects the exciton population distribution in $k$-space. When the excitation is off-phonon resonance, it generates excitons with a finite kinetic energy, resulting in excitons with a wide steady-state $k$-space distribution (see Fig. S4a). A large number of excitons have high energies, and cannot be efficiently scattered into the light cone for radiative recombination due to energy conservation. On the other hand, when the excitation is on-phonon resonance, it generates excitons with close to zero kinetic energy. The subsequent exciton steady-state $k$-space distribution is narrower (Fig. S4b). When the excitation laser is tuned across phonon resonances, the variation of exciton distribution in $k$-space leads to the observed oscillations in PLE luminescence spectra.

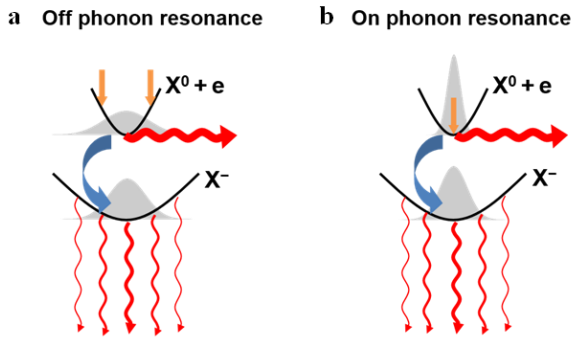

**Fig. S4** Exciton and trion steady-state $k$-space distributions for **a** on- and **b** off-phonon-resonance excitation. The upper parabola represents the dispersion curve of an exciton together with a free electron ($X^0$ + e), while the lower that of a trion ($X^-$). Orange arrows indicate laser generated exciton energies following phonon emission (see Fig. 5b & c in the main text), while blue curved arrows illustrate the relaxation process from a neutral exciton to a trion. Red wavy arrows represent the radiative decay process, whose thickness corresponds to the wavevector-dependent decay rate. Only excitons lying in the vicinity of valley extremum can radiatively recombine, while all trions can radiatively recombine with a slowly changing decay rate. Shaded Gaussian curves correspond to the exciton and trion $k$-space distributions.

Most trions are formed via the relaxation of excitons[7]. This additional step can partially reduce the difference of trion $k$-space distributions between exciting on- and off-phonon resonance. Meanwhile, trions in a much larger momentum range can radiatively decay with a slowly changing decay rate (Fig. S4). This further diminishes the variation of trion emission as a function of excitation energy.

## V. Data from a second monolayer MoSe$_2$ sample.

Data from a second monolayer MoSe$_2$ sample (Fig. S5a) exfoliated from a different bulk crystal is presented in Fig. S5b – g. In this sample, the resonances for both X$^-$ and X$^0$ are blue-shifted by about 3 meV relative to the first sample. Nonetheless, the phonon resonances are identical (within the uncertainty of 1 meV) to those found in the first sample, as can be seen in the vertical line cut at the X$^0$ resonance shown in Fig. S5c. Here, we have included additional Raman peak assignments in addition to the LA(M) overtones where space allows. These assignments are provisional, especially for higher order modes, because multiple phonon combinations may give rise to the same energy.

The emission lineshapes of X$^0$ and X$^-$ in the second monolayer MoSe$_2$ is different from those in the first in the sense that they are slightly asymmetric. As a result, fitting using two Voigt profiles as is done in the previous case becomes inadequate here. Nonetheless, we found that excellent fits can be produced by adding two Gaussian curves whose peak positions and widths are constant, i.e., independent of the excitation energy, as shown in Fig. S5d. Their peak positions are each about 10 meV lower than X$^0$ and X$^-$ resonances. Given their large linewidths of about 7.5 meV, they are unlike to be originating from defect states. It is more likely that their presence is due to contamination or inhomogeneity as a result of lattice distortion. Despite this, HWHM of X$^0$ emission, $\gamma_0$, extracted from the lineshape analysis shows oscillations with a period close to the LA(M) phonon mode (Fig. S5e), as is the case in the first sample. However, oscillations are also noticeable in the inhomogeneous broadening width, $\sigma$, as shown in the inset of Fig S5e, although the amplitude of oscillations (about 0.15 meV) is much smaller than that of $\gamma_0$. This is likely an artifact of the fit due to the weaker X$^0$ emission intensity and the more pronounced Raman scattering components present in the second sample. Regardless, the oscillatory behavior of the homogeneous linewidth and of the peaks position of X$^0$ (Fig. S5f) qualitatively agrees with results obtained from the first sample.

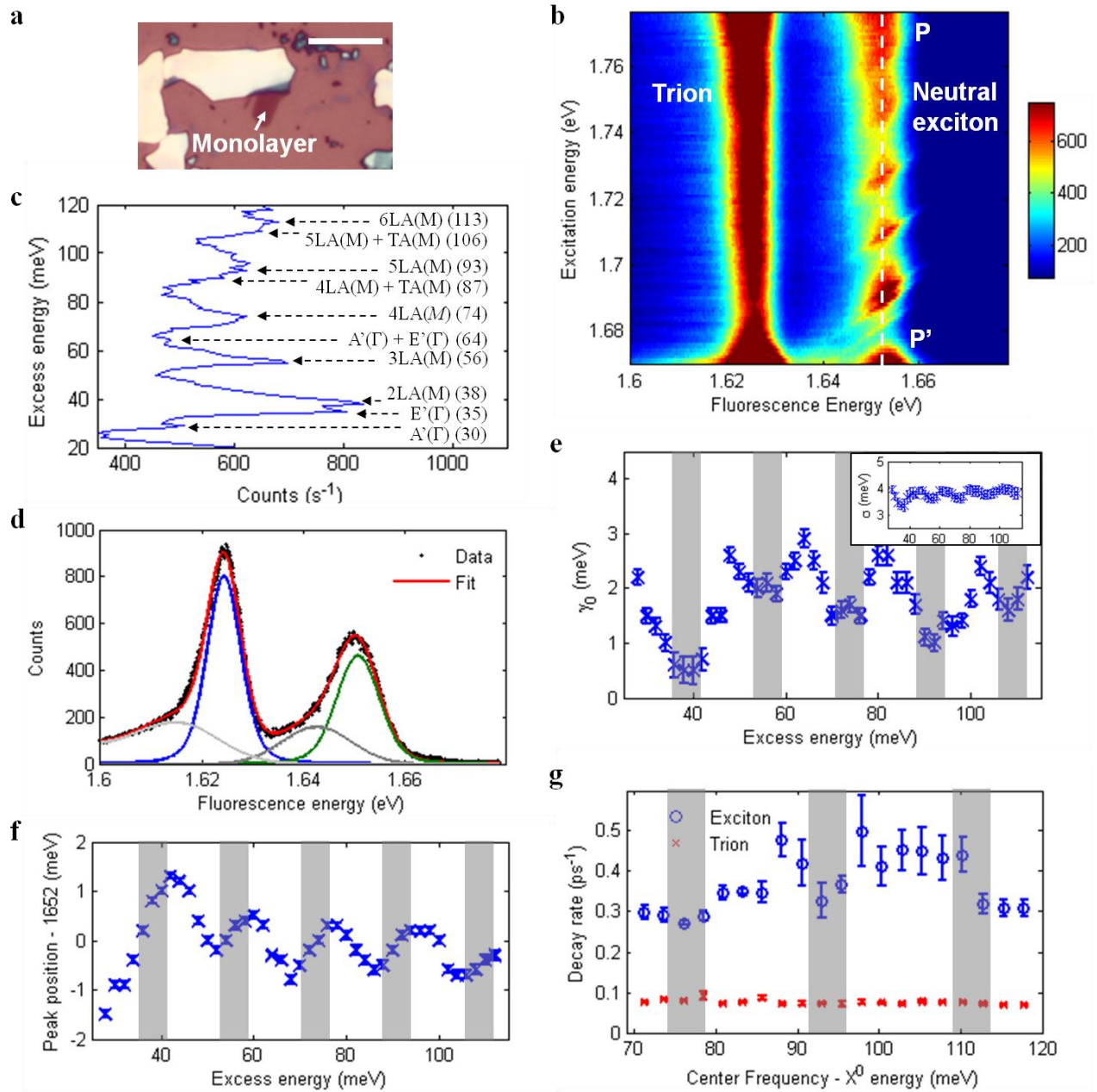

**Fig. S5** **a** Optical micrograph of a second monolayer $MoSe_2$. Scale bar: 10 μm. **b** PLE spectra of the second sample, showing neutral exciton and trion emission centered at 1.653 eV and 1.626 eV, respectively. Colour bar: counts per second. **c** Vertical line cut P – P' of the PLE spectra shown in **b** with selected phonon peaks marked by arrows. **d** An example of the PL emission spectra with fit. Solid black circles are data points while the solid red line is the fit constructed from the following components: Voigt profiles for trion (solid blue line) and neutral exciton (solid greed line), and the Gaussian background (solid dark and light grey). **e** Best-fit homogeneous linewidth, $\gamma_0$, of the $X^0$ resonance as a function of excess energy extracted from lineshape analysis of the PLE intensity map in **b**. Inset: Width of the inhomogeneous (Gaussian) broadening, $\sigma$. **f** Peak position of the $X^0$ resonance as a function of excess energy. Error bars in both **e** and **f** represent 99% confidence intervals of the fits, while shaded regions correspond to regions of enhanced luminescence seen in **b**. **g** Excitation frequency-dependent decay rates, $\gamma$'s, of $X^-$ (crosses) and $X^0$ (circles) extracted by fitting the exponential decay time of the emission.

## VI. Raman spectrum of a third monolayer MoSe$_2$ sample

Raman spectrum of a third monolayer MoSe$_2$ sample is shown in Fig. S6 below. The spectrum is taken at room temperature, with an excitation laser wavelength and intensity of 514 nm and 2 mW, respectively. The resonances observed agree with the spectra obtained by other groups[9–11], as well as our PLE data shown in Fig. 2c in the main text. The linewidths of the prominent peaks, e. g. $A_1'$ and $E'$, are found to range from 0.5 to 1 meV, matching those in the PLE spectra.

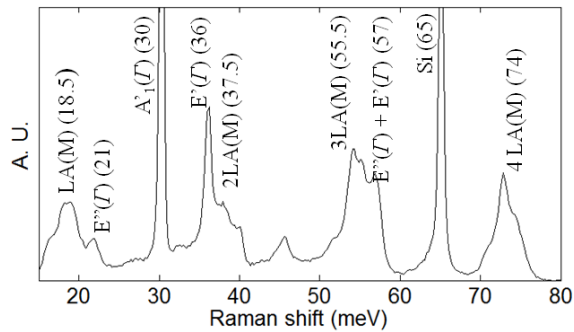

**Fig. S6** Raman spectrum of a third monolayer MoSe$_2$ sample. Prominent peaks are labeled with corresponding phonon modes, along with the Raman shifts (in units of meV) in parentheses.